\documentstyle[preprint,amssymb,aps]{revtex}

\begin{document}

\title{Inhomogeneity of dusty crystals and plasma diagnostics}
\author{L.I. Podloubny$^{(1)}$, P.P.J.M. Schram$^{(2)}$ and S.A. Trigger$^{(1,2)}$}
\address{$^{(1)}$ Institute for High Temperatures, Russian Academy of Sciences,\\
Izhorskaya St., 13/19, 127412, Moscow, Russia.\\
$^{(2)}$ Eindhoven University of Technology P.O. Box 513, 5600 MB Eindhoven,
\\
The Netherlands}
\maketitle

\begin{abstract}
Real dusty crystals are inhomogeneous due to the presence of external
forces. We suggest approximations for calculations of different types
of inhomogeneous DC (chain and DC with a few slabs) in the equilibrium state.
The results are in a good agreement with experimental results and can be
used as an effective diagnostic method for many dusty systems.

PACS number(s): 52.25.-b, 64.70.-p, 94.10. Nh
\end{abstract}

\section{Introduction}

Formation of dust crystals (DC) takes place in a vertical electric field
of the sheath, the gravitational field and a horizontal electrical field. The
external field, acting in vertical and horizontal traps, stabilizes the
3-dimensional DC of finite size and linear chains of $d$-ions (horizontal
traps for confinement of one-dimensional DC are used in \cite{Peters,Homm}).
The pressure of the boundaries and the external field violate
the translational invariance and lead to a dependence of the distances
between nearest neighbors in the lattice of dust particles on the position of
the particles (see Fig.\ref{fig1} and Fig.\ref{fig2}). Therefore macroscopic inhomogeneity in a 
lattice is a new phenomenon not present in the usual infinite (very large) 
crystal.

Even in the approximation of central force for interparticle interaction
between $d$-ions, DC possess a layered structure (the layered structure
of usual atomic crystals, as graphite, is connected with the anisotropy of
the interparticle interaction).

The vertical and horizontal distances between nearest neighbours (lattice
``constants'' $R_{\parallel }$ and $R_{\perp }$) are in general different
functions of position in different directions from the center of the
crystal (center of inertia). Deformation of DC in the fields of the traps
depends on its characteristics and on the plasma parameters. Therefore
the electrostrictional response of $d$-ion systems on a static external
disturbance can be used as a diagnostic tool for DC and the surrounding
plasma. In particular the charge $Q$ of $d$-ions, the screening length $R_{D},$
the concentration of the small ions and the electric field in the sheath
can be determined. In the present paper the possibility to use the
inhomogeneity of DC for plasma diagnostics is considered theoretically.

Recently dusty plasma diagnostics appear on basis of investigations of
the dispersion curves $\omega (k)$  for $d$-ion sound \cite{Pieper}
and properties of forced oscillations of linear $d$-ion's chains in an
electric field \cite{Peters} and under the action of laser impulses \cite{Homm}.
The static diagnostic, suggested in this paper is simpler for the
theoretical description and experimental realization than the dynamic
sounding considered in \cite{Peters,Homm,Pieper}.

For the description of a lattice configuration of $N$ $d$-ions in a
state of deformation under action of external gravitational and electric
forces $\overline{f_{n}}=-\nabla V_{n}$ and interparticles forces $\overline{
F_{n}}=-\nabla U_{n}$ we will use the balance equations. Here $V_{n}$ is the
potential energy of the $d$-ion with number $n$, $U_{n}$ is the potential
energy of interaction between the $d$-ion with number $n$ and all other ones. We
do not take into account the force connected with momentum transfer from
the small ions to the $d$-ions. This force very often can be omitted,
because in the case when it is essential, the $d$-ions can be found not only
below the sheath, but also on top of it, which is not observed in the
experiment discussed below.

We will use the simple approximation of nearest neighbours for the
description of interparticle interaction. This approximation apparently
gives a good picture of the inhomogeneity of DC under the action of
external forces and with a screening length $R_{D}\sim R_{\perp },R_{\parallel
} $. We also will neglect a possible dependence of the $d$-ion charge $Q$
on the location in the inhomogeneous DC (w.r.t. $d$-ion density).
Therefore we suggest $Q=const$ in our considerations.

\section{Equations of static equilibrium}

For the case of inhomogeneous three-dimensional DC we will use a simple
quasi-one-dimensional model of DC, in which the layer lattice with a real
potential is changed into a one-dimensional vertical chain of particles. The
effective potential for this model can be calculated by integration of
the interaction with the nearest layer with distributed charge $\sigma =Q/S_{0}$\,\,($S_{0}$ is the surface for a $d$-ion in horizontal direction)
\begin{equation}
\left\langle U(r)\right\rangle _{xy}=\frac{2\pi }{S_{0}}\int\limits_{0}^{\infty }d\rho \rho U\left( \sqrt{z^{2}+\rho ^{2}}\right)  \label{1}
\end{equation}

For the Debye-Hueckel interaction and simple hexagonal lattice the potential (\ref{1}) has the form

\begin{equation}
U(z)=\left\langle \frac{Q^{2}}{2}e^{-\varkappa r}\right\rangle
_{x,y}=U_{0}e^{-\varkappa z},\quad \varkappa =\frac{1}{R_{D}},\quad
U_{0}=2\pi \frac{Q\sigma }{\varkappa },\quad \sigma =\frac{2Q}{\sqrt{3}
R_{\perp }^{2}}  \label{2}
\end{equation}

This model permits to calculate the dependence of the distances between the
nearest slabs $R_{n}(z)$ as a function of height.

In the general case of pair interaction between the $d$-ions in the external
electric and gravitational fields of the sheath the potential energy can be
written in the form

\begin{equation}
U+V=\sum_{k=1}^{N-1}U_{k}+\sum_{k=1}^{N}V_{k},\quad U_{k}=U(R_{k}),\quad
V_{k}=V(z_{k}),\quad R_{k}=z_{k+1}-z_{k}  \label{3}
\end{equation}

Here we take into account only interaction between neighbouring particles.
The potential energy for a horizontal chain of $N$ interacting $d$-ions in the
electric field of the trap, has an analogous form and stabilizes this chain in
the $x$-direction $(z_{k}\rightarrow x_{k})$.
The conditions of balance of external and internal forces lead to a system
of equations which determines the configuration of the $d$-ions:

\begin{equation}
\left\{
\begin{array}{l}
U_{k}^{\prime }-U_{k+1}^{\prime }+V_{k+1}^{\prime }=0,\quad
k=1,2,3,...N-2,\quad U_{k}^{\prime }=%
{\displaystyle {dU \over dR_{k}}}%
\quad V_{k}^{\prime }=%
{\displaystyle {dV \over dz_{k}}}%
, \\
-U_{1}^{\prime }+V_{1}^{\prime }=0, \\
U_{N-1}^{\prime }+V_{N}^{\prime }=0.
\end{array}
\right.  \label{4}
\end{equation}

Summation of the left parts of these equations leads to the obvious condition
of zero sum of the external fields: $\sum\limits_{k=1}^{N}V_{k}^{\prime }=0$.

For the stabilization of horizontal chains an external field in the form of
a parabolic well in the chain direction has been used in \cite{Peters,Homm}.

\begin{equation}
V_{k}=\frac{1}{2}m\omega _{0}^{2}(x_{n}-X_{0})^{2},\quad X_{0}=\frac{1}{N}%
\sum_{k=1}^{N}x_{k}  \label{5}
\end{equation}

Here $X_{0}$ is the center of inertia for a chain and $\omega _{0}$ is a
parameter, which determines the shape of the pit. According to \cite{Somm}
the vertical electric field in a sheath changes linearly with the height. This
dependence is realized approximately in the regions not too close to the lower
electrode and the border of the presheath: the quadratic approximation for
the potential $\varphi (z)$ in the plasma layer is also used in
\cite{Melzer,Schw} for the analysis of the equations of motion of DC . Therefore in the case
of a vertical potential well we use in eq.(4) the expansion

\begin{equation}
\begin{array}{l}
V(z_{k})=mgz_{k}+Q\varphi _{0}+Q\varphi _{0}^{\prime }(z_{k}-X_{0})+\frac{1}{%
2}\omega _{0}^{2}(z_{k}-X_{0})^{2}, \\
\varphi _{0}=\varphi (X_{0}),\quad \varphi _{0}^{\prime }=\varphi ^{\prime
}(X_{0}),\quad m\omega _{0}^{2}=Q\varphi ^{\prime \prime }(X_{0}).
\end{array}
\label{6}
\end{equation}

The parabolic approximation for the vertical electric field is reasonable
for the case of sufficiently thin DC . To estimate the maximal thickness $\ell
=z_{N}-z_{1}=2(X_{0}-z_{1})$ of DC, for which this approximation is true,
let us consider $\varphi (z)=\varphi (0)\exp (-z/R_{D})$ and use the
condition

\[\frac{1}{3}\left| Q\varphi _{0}^{\prime \prime \prime }\right|
(X_{0}-z_{1})^{2}\sim \frac{1}{3}\left| Q\varphi _{0}^{\prime }\right|
\left( \frac{\ell }{2R_{D}}\right) ^{2}<Q\varphi _{0}^{\prime \prime
}(X_{0}-z_{1})\sim \left| Q\varphi _{0}^{\prime }\right| \frac{\ell }{2R_{D}}\]

Then the necessary inequality is $\ell <6R_{D}$ which is usually satisfied
(see, for example, \cite{Pieper,Melzer,Schw}). The linear terms in Eq.(\ref
{6}) are really absent because of the condition of zero total external
forces:

\begin{equation}
mg+Q\varphi ^{\prime }(X_{0})=0.  \label{mg}
\end{equation}

This condition determines the position of the center of inertia for the
system of levitated $d$-ions.

By use of the parabolic approximation (\ref{6}) in the balance equations (\ref
{4}) and subtracting from each equation the previous one, we find

\begin{equation}
\left\{
\begin{array}{l}
2U_{k}^{\prime }-U_{k+1}^{\prime }-U_{k-1}^{\prime }+m\omega
_{0}^{2}R_{k}=0,\quad k=2,3,...N-2, \\
2U_{1}^{\prime }-U_{2}^{\prime }+m\omega _{0}^{2}R_{1}=0, \\
2U_{N-1}^{\prime }-U_{N-2}^{\prime }+m\omega _{0}^{2}R_{N-1}=0.
\end{array}
\right.  \label{7}
\end{equation}

As follows from eq.(\ref{7}) the intervals $R_{k}$ are symmetric with
respect to the center:

\[
R_{1}=R_{N-1},\quad R_{2}=R_{N-2},\quad ...,\quad R_{k}=R_{N-k},\quad ...
\]

\section{Structure of DC with an attractive (for large distances) and with
a purely repulsive potential}

According to eqs.(\ref{4}), (\ref{7}) for isolated systems of $d$-ions ($
V_{k}^{\prime }=0$),  there are two different possibilities when external
fields are absent.

If the pair interaction between $d$-ions is a nonmonotonic function and
leads to repulsion at small distances and to attraction at large
distances, then the solution of eq.(\ref{4}) reads

\begin{equation}
U_{1}^{\prime }=U_{2}^{\prime }=...=U_{N-1}^{\prime }=0  \label{8}
\end{equation}.

This solution describes a homogeneous chain of $N$ $d$-ions with equal
distances between nearest neighbours $R_{1}=R_{2}=...=R_{N-1}=R_{0}$. The potential energy  has a minimum for this
state. In this case the weakly inhomogeneous configurations of $d$-ions with
external force $V_{k}^{\prime }\neq 0$ can be described on basis of small
deformations $|R_{0}-R_{k}|\ll R_{0}$. Then we use the expansion

\begin{equation}
U(R_{k})=U_{0}+\frac{m\Omega ^{2}}{2}(R_{k}-R_{0})^{2},\quad m\Omega
^{2}=U^{\prime \prime }(R_{0})  \label{U(Rk)}
\end{equation}

If the pair interaction $U(R_{k})$ has a monotonic purely repulsive form,
the $d$-ions of an isolated system are unstable and, according to (\ref{8})
all $R_{k}$ are infinite. In this case stabilization of the system in a weak
external field playing the role of a trap, leads also to a slightly
inhomogeneous state, in which the deviations of the intervals from the
average are small,

\begin{equation}
R_{0}=\frac{1}{N-1}\sum_{k=1}^{N-1}R_{k},\quad \left| R_{0}-R_{k}\right| \ll
R_{0}  \label{10}
\end{equation}

In this case the alternative quadratic expansion of the energy for $d-d$
interactions has the form

\begin{equation}
\begin{array}{l}
\mathop{\displaystyle \sum }%
\limits_{k=1}^{N-1}U_{k}=(N-1)U_{0}+U_{0}^{\prime }%
\mathop{\displaystyle \sum }%
\limits_{k=1}^{N-1}(R_{k}-R_{0})+\frac{1}{2}m\Omega ^{2}%
\mathop{\displaystyle \sum }%
\limits_{k=1}^{N-1}(R_{k}-R_{0})^{2}= \\
\quad \quad \quad (N-1)U(R_{0})-(N-1)U_{0}^{\prime }R_{0}+U_{0}^{\prime
}(z_{N}-z_{1})+\frac{1}{2}m\Omega ^{2}%
\mathop{\displaystyle \sum }%
\limits_{k=1}^{N-1}(R_{k}-R_{0})^{2}, \\
U_{0}=U(R_{0}),\quad U_{0}^{\prime }=%
{\displaystyle {dU(R_{0}) \over d(R_{0})}}%
,\quad m\Omega ^{2}=%
{\displaystyle {d^{2}U(R_{0}) \over dR_{0}^{2}}}%
,\quad
\mathop{\displaystyle \sum }%
\limits_{k=1}^{N-1}R_{k}=z_{N}-z_{1}.
\end{array}
\label{11}
\end{equation}

For a potential with a well $U^{\prime }(R_{0})=0$ the expansions (\ref
{U(Rk)}) and (\ref{11})  coincide, therefore small deformations $
s_{k}=R_{0}-R_{k}$ of the system in an external field can then be described
by the general equations of force balance:

\begin{equation}
\left\{
\begin{array}{l}
2\cosh t\cdot s_{k}-s_{k+1}-s_{k-1}-%
{\displaystyle {\omega _{0}^{2} \over \Omega ^{2}}}%
R_{0}=0,\quad \cosh t=1+%
{\displaystyle {\omega _{0}^{2} \over 2\Omega ^{2}}}%
,\quad k=2,3,...,N-2 \\
2\cosh t\cdot s_{1}-s_{2}-%
{\displaystyle {\omega _{0}^{2} \over \Omega ^{2}}}%
R_{0}-%
{\displaystyle {U_{0}^{\prime } \over m\Omega ^{2}}}%
=0, \\
2\cosh t\cdot s_{N-1}-s_{N-2}-%
{\displaystyle {\omega _{0}^{2} \over \Omega ^{2}}}%
R_{0}-%
{\displaystyle {U_{0}^{\prime } \over m\Omega ^{2}}}%
=0.
\end{array}
\right.  \label{12}
\end{equation}

Here for purely repulsive interaction $R_{0}$ is the average. For the
case with attraction $U_{0}^{\prime }=0$ and $R_{0}$ is the equilibrium
distance in the isolated system of $d$-ions.

\section{Solutions and numerical results}

A general solution of the equations in finite differences (\ref{12}) can be
obtained in the form

\begin{equation}
s_{k}=R_{0}-R_{k}=R_{0}+Ae^{kt}+Be^{-kt}.  \label{13}
\end{equation}

Taking into account the symmetry of the system $s_{k}=s_{N-k}$, the connection
between the coefficients $B=Ae^{Nt}$ can be found. The coefficient $A$ can
be found from the boundary condition for $k=1$ (or for $k=N-1$). Finally for
the interval number $k$ and purely repulsive potential we find

\begin{equation}
R_{k}=\left( R_{0}-%
{\displaystyle {U^{\prime }(R_{0}) \over m\Omega ^{2}}}%
\right) \frac{\cosh \left( \frac{N}{2}-k\right) t}{\cosh \frac{Nt}{2}}%
=\left( R_{0}-%
{\displaystyle {U^{\prime }(R_{0}) \over m\Omega ^{2}}}%
\right) \cdot \frac{C_{k-1}^{\prime }(\cosh t)+C_{N-k-1}^{\prime }(\cosh t)}{%
C_{N-1}^{\prime }(\cosh t)}  \label{Rk}
\end{equation}

Here $C_{n}^{\prime }(x)$ are the Gegenbauer polynomials. For the case of
interaction with attraction $U^{\prime }(R_{0})=0$, the intervals $R_{k}$
have the form:

\begin{equation}
R_{k}=R_{0}\frac{\cosh \left( \frac{N}{2}-k\right) t}{\cosh \frac{N}{2}t}.
\label{Rk2}
\end{equation}

Therefore in the parabolic trap formed by the external forces, a chain of $d$
-ions is compressed symmetrically w.r.t.\, the center of inertia, and the
central regions more strongly than the ones outwards $R_{1}>R_{2}>...$ For the
resulting electrostrictional reduction of the length $\ell $ of an entire
chain it follows from Eq.(\ref{Rk2}) that ($R_{0}$ is the equilibrium
distance in a homogeneous chain)

\begin{equation}
\ell =\sum_{k=1}^{N-1}R_{k}=2R_{0}\frac{\cosh
{\textstyle {t \over 2}}%
\sinh \frac{N-1}{2}t}{\sinh t\cdot \cosh \frac{Nt}{2}}<(N-1)R_{0}
\label{ell12}
\end{equation}

For sufficiently long ($N\gg 1$) horizontal chains and for (in vertical
direction) quasi-one-dimensional  dusty crystals the profile distributions of
charge density and mass and thereby the ``constants'' of the elastic
forces can be obtained in the approximation of continuous media by use of
eqs.(\ref{Rk2}), (\ref{ell12}). The surface density of charge is proportional to
to the mass density and therefore there is balance of the external volume
electric and gravitational forces in each point of a horizontal plane at
fixed height. This means that even inhomogeneous planes (Fig.\ref{fig1}) and horizontal 
chains (Fig.\ref{fig2}), which are more dense in the center, are not suspended in the 
central part of the dusty system, where the density is higher. Enlargement of 
the density in the center of horizontal crystalline planes is observed in the 
experiments \cite{Goree}, but quantitative measurements are unknown to us. 
Parallel to oscillation and wave measurements in horizontal chains, the 
equilibrium positions of $d$-ions have also been determined in the electric 
field of a horizontal trap \cite{Peters,Homm,Melzer2}. According to the data 
of these papers for the case $N=12$ the ratios of the intervals between 
neighbouring $d$-ions in the direction of the center are 
$R_{1}:R_{2}:...R_{6}=1.44:1.22:1.11:1.05:1.01:1.00$. These results are 
reasonably described by our formula (\ref{Rk}), in which for $t=0.18$ 
(and correspondingly $\omega _{0}=0.2\Omega $) these ratios are 
$1.43:1.27:1.15:1.07:1.02:1.00$.

The experimental data for the other half of the chain $R_{6}:R_{7}:...R_{11}=
1:1.01:1.01:1.08:1.20:1.32$  agree less with our theory for the (with respect
to the center) symmetric chain and they are  essentially different from the
experimental data for the first half of the chain. We think that this
asymmetry is a consequence of the asymmetric and not exactly parabolic
$V(x)\approx \frac{1}{2}m\omega _{0}^{2}x^{2}$ shape of the external electric
field (here $x$ is the distance from the center of the chain). According to
\cite{Melzer2} $m\omega _{0}^{2}=2.55\cdot
10^{-11}$ kg$\cdot $s$^{-2}$, $m=6.73\cdot 10^{-13}$ kg. Using the data on
the equilibrium configuration $R_{n}$ and the parameter of the trap
$\omega _{0}=6.15 $ s$^{-1}$ we find the important characteristic of $d-d$
interaction $\Omega =5\omega _{0}=30.7$ s$^{-1}$.

For the chain with $N=4$ the experimental data, according to \cite{Peters,Melzer2},
give $\omega _{0}=6.25$ s$^{-1}$ and $R_{1}=1989$ $\mu $m,
$R_{2}=1910$ $\mu $m, $R_{3}=2031$ $\mu $m. The average interval $R_{\perp
}=1960$ $\mu $m for the case $N=4$ is twice as large as $R_{\perp }=10^{3}$
$\mu$m for the chain with $N=12$. This is probably connected with the higher
charges (almost three times) of the $d$-ions in \cite{Peters} and therefore
with the stronger repulsion between them at the partially same compressing
external field of the horizontal trap. For the conditions of the experiments \cite{Peters}
the asymmetry of the external field, connected with the nonquadratic form
of the potential $V(x)$ is still stronger than in \cite{Homm} and this was
the reason to use for the bordering intervals the expression $\left\langle
R_{1}\right\rangle =(R_{1}+R_{3})/2=2010$ $\mu $m. Then according to (\ref
{U(Rk)}) we have $\left\langle R_{1}\right\rangle /R_{2}=1+\omega
_{0}^{2}/2\Omega ^{2}=1.05$ and $\Omega =3.16\omega _{0}=19.7$ s$^{-1}$.

It is necessary to emphasize that all the results for the case $N=4$ and $N=12$
are applicable for both cases: purely repulsive $d-d$ interaction and
$d-d$ interaction with an attractive part, because, as follows from
eqs.(\ref{Rk}) and (\ref{Rk2}), the ratios of intervals $R_{k}$ are the same
in these cases.

The known experimental data on equilibrium intervals $R_{\parallel }$
between neighbouring ions in vertical traps concern only dust crystals with
two horizontal crystalline planes ($N=2$) and have been obtained in
\cite{Melzer,Schw}. In \cite{Pieper} a dust crystal with $N=3$ has been
investigated but the thickness of the crystal was not measured.

According to \cite{Melzer,Schw} the ratio $(R_{0}-R_{\parallel })/R_{0}=0.2$
and does not depend on the ion's mass.

In \cite{Melzer} experiments are reported with dust crystals, formed by
$d$-ions with radii 4.7 $\mu $m and 2.4 $\mu $ which leads to a difference
of gravitational force proportional to $m_{1}/m_{2}\simeq 8$. The position
of the center of inertia $X_{0}$ of the dust crystal must be considerably
changed in this case: a lighter crystal will shift over a distance $\sim R_{D}$,
as follows from Eq.(\ref{mg}). A measurement of this effect was
not reported in \cite{Melzer}.

According to (\ref{Rk2})

\begin{equation}
1-\frac{R_{\parallel }}{R_{0}}=\frac{\omega _{0}^{2}}{\omega
_{0}^{2}+2\Omega ^{2}}=0.2  \label{17}
\end{equation}

and therefore $\Omega =\sqrt{2}\omega _{0}$. In contrast with \cite{Peters,Homm}
the parameter $\omega _{0}^{2}=\frac{1}{m}V_{0}^{\prime \prime
}(x)$ for the vertical electric field in a sheath is here unknown. It can be
determined only on basis of knowledge of the interaction potential between the
$d$-ions, via the parameter $\Omega ^{2}=\frac{1}{m}U^{\prime \prime}(R_{\parallel })$.

Purely repulsive interaction (\ref{1}) leads to another result. For $N=2$
the exact system of balance eqs.(\ref{3}) has the form

\begin{equation}
\left\{
\begin{array}{l}
-U^{\prime }(R_{\parallel })+V^{\prime }(z_{1})=0,\quad R_{\parallel
}=z_{2}-z_{1},\quad X_{0}=%
{\displaystyle {z_{1}+z_{2} \over 2}}%
, \\
U^{\prime }(R_{\parallel })+V^{\prime }(z_{2})=0,\quad z_{2,1}=X_{0}\pm \frac{%
1}{2}R_{\parallel }.
\end{array}
\right.  \label{18}
\end{equation}

In this case a more general model for the external potential (\ref{6}) than
the linear one can be used for the description of the electric field
in a sheath. Let us take

\begin{equation}
E(X_{0}\pm \frac{1}{2}R_{\parallel })=E(X_{0})e^{\pm \frac{1}{2}\varkappa
R_{\parallel }}.  \label{19}
\end{equation}

For the position of the center of inertia we have

\begin{equation}
mg=QE(X_{0})\cosh \frac{1}{2}\varkappa R_{\parallel },  \label{2mg}
\end{equation}

and according to (\ref{18}) with $U(R_{\parallel })$ taken from (\ref{2}) we
find

\begin{equation}
V^{\prime }(z_{2})-V^{\prime }(z_{1})=2QE(X_{0})\sinh \frac{1}{2}\varkappa
R_{\parallel }=2\frac{4\pi Q^{2}}{\sqrt{3}R_{\perp }^{2}}e^{-\varkappa
R_{\parallel }}.  \label{21}
\end{equation}

By eliminating  $E(X_{0})$ from these equations we finally find

\begin{equation}
\tanh \frac{\varkappa R_{\parallel }}{2}=\alpha e^{-\varkappa R_{\parallel
}},\quad \alpha =\frac{4\pi Q^{2}}{\sqrt{3}R_{\perp }^{2}mg}.  \label{22}
\end{equation}

Using the experimental data \cite{Melzer} for $m$, $Q$, $R_{\perp }=450\,\mu
$m, $R_{\parallel }=360$ $\mu $m we estimate the Debye radius $R_{D}=973$ $
\mu $m. In the case of a dust crystal with a lower $m$, $Q$ and $R_{\perp
}=350\,\mu $m, $R_{\parallel }=280\,\mu $m (see also \cite{Melzer}) we find $
R_{D}=933\,\mu $m. Therefore the Debye length $R_{D}$ is approximately of the
order of the interval between nearest $d$-ions, which is in agreement
with the estimates of \cite{Schw}.

For the electric field in a sheath we find from eq. (\ref{2mg})
$E(X_{0})=2.84\cdot 10^{3}$\thinspace V$\cdot $m$^{-1}$ and $E(X_{0}+\Delta
)=1.99\cdot 10^{3}~$V$\cdot $m$^{-1}$ for the case of $d$-ions with
radii 4.7 $\mu $m and 2.4 $\mu $m.

Let us neglect small changes of Debye radius $R_{D}$ and take

\begin{equation}
\frac{E(X_{0})}{E(X_{0}+\Delta )}=\exp \left( \frac{\Delta }{\left\langle
R_{D}\right\rangle }\right) =1.42,\quad \left\langle R_{D}\right\rangle
=950\,\mu \text{m}.  \label{RD950}
\end{equation}

Then we obtain for the shift upwards $\Delta $ of the lighter crystal

\begin{equation}
\Delta =0.35\left\langle R_{D}\right\rangle =332\,\mu \text{m}.  \label{Delta}
\end{equation}

The moving of a dust crystal inside the sheath can be observed by different
microgravity experiments (see some discussion for example in \cite{Morfill}).

We suggest here some experiments in which the properties of DC can be
studied under conditions of microgravity and even changing gravity.

One of these experiments (under terrestrial conditions) can be performed in
a horizontal discharge, where in the horizontal direction there is only an
electric force and momentum transfer from the small ions to the dust
particles. For such experiment the latter force can be very essential in
contrast to the conditions considered in this paper.

The second group of experiments is connected with the effective gravity
created in space stations by rotation of dusty plasma. If $h$ and $g_{
\text{eff}}$ are the distance from the axis of rotation to the negative
electrode and the acceleration of the center of inertia for the dusty system
respectively, the obvious connection is given by

\begin{equation}
g_{\text{eff}}=\omega ^{2}(h-X_{0}).  \label{geff}
\end{equation}

For $g_{\text{eff}}=g$ and $h=1$ m (rotation of the container inside the
space station or rocket) or $h=10$ m (rotation of the space station as a
whole) we find $\omega =3$ s$^{-1}$ and $\omega =1$ s$^{-1}$ respectively,
which are conditions of weakly inhomogeneous ($h\gg R_{k}$) artificial
gravitational field where our results, obtained above, are applicable.
Measuring the dependence $X_{0}=X_{0}(\omega )$ would permit to investigate
the profile of the electric field in a sheath and other characteristics of the
dusty system and plasma. Of special interest is the investigation of the
deformation of DC in an essentially inhomogeneous rotation field
($h-X_{0}\sim 0.05$ m and $\omega \sim 15$ s$^{-1}$). A detailed
consideration of such experiment will be given in a separate paper.

In the case of a dust crystal with three horizontal crystalline planes the
static equilibrium is described by the system of eqs.(\ref{4}) with $N=3$.
In the approximation for the electric fields used before the coordinate
of the average $d$-ions $z_{2}$ and the value $E_{0}$ can be eliminated on
basis of balance of external fields:

\begin{equation}
3mg=QE_{0}\sum_{n=1}^{3}e^{-\varkappa z_{n}}=QE_{0}e^{-\varkappa
z_{2}}(1+e^{\varkappa R_{1}}+e^{-\varkappa R_{2}}).  \label{3mg}
\end{equation}

For the system of equations determining the vertical intervals $R_{1}$
and $R_{2}$, we obtain

\begin{equation}
\begin{array}{l}
\frac{\alpha }{3}(e^{-\varkappa R_{1}}-2e^{-\varkappa R_{2}})+
{\displaystyle {1-e^{-\varkappa R_{2}} \over 1+e^{\varkappa R_{1}}+e^{-\varkappa R_{2}}}}%
=0, \\
\frac{\alpha }{3}(e^{-\varkappa R_{2}}-2e^{-\varkappa R_{1}})+
{\displaystyle {e^{\varkappa R_{1}}-1 \over 1+e^{\varkappa R_{1}}+e^{-\varkappa R_{2}}}}%
=0.
\end{array}
\label{24}
\end{equation}

Even for the highest pressure of neutrals in \cite{Pieper}, $p=300$ mTorr ($Q=7.2\cdot
10^{3}$e, $R_{\perp }=0.28$\thinspace mm, $\varkappa R_{\perp }=0.61$) the
parameter $\alpha /3=0.053\ll 1$. Suggesting $R_{1}=R_{2}=R_{\parallel }$
and $\varkappa R_{\parallel }\ll 1$ it follows from eq.(\ref{24}) that

\begin{equation}
\varkappa R_{\parallel }\approx \frac{\alpha }{1+\alpha }=0.14.  \label{25}
\end{equation}

Let us emphasize that the vertical compression is symmetric $(R_{1}=R_{2})$
with respect to the central plane only for $\varkappa R_{\parallel }\ll 1$. In
contrast to the approximate equations (\ref{7}) for the parabolic wells, the
exact equations (\ref{24}) are not symmetric for the interchange
$R_{1}\rightleftarrows R_{2}$.

From eq.(\ref{3mg})-(\ref{25}) with the Debye radius given in
\cite{Pieper} it follows
that vertical compression is important:

\begin{equation}
\frac{R_{\perp }-R_{\parallel }}{R_{\perp }}=0.77.  \label{26}
\end{equation}

In the framework of the quadratic approximation for the potential energy of
the system with $N=3$ we find according to eqs.(\ref{Rk})-(\ref{Rk2})

\begin{equation}
\frac{R_{\perp }-R_{\parallel }}{R_{\perp }}=\frac{\omega _{0}^{2}}{\omega
_{0}^{2}+\Omega ^{2}}.  \label{27}
\end{equation}

Unfortunately the vertical interval $R_{\parallel }$ has not been measured
in \cite{Pieper}. The experimental data obtained in \cite{Pieper} are not sufficient to choose
which variant is preferable for purely repulsive interaction or interaction
with an attractive part: the quadratic model or the more exact description
(\ref{3mg})-(\ref{24}).

\section{Conclusions.}

The method of dusty plasma diagnostics discussed above and based on an
analysis of the inhomogeneity of the linear structures of $d$-ions, seems
very attractive. In contrast to the situation in the usual sound method, the
$d$-ions of a  small dust crystal or a linear dust chain have additional
degrees of freedom. It gives the possibility to extract additional information
from the static response (change of the equilibrium distances between the
$d$-ions) or the dynamic response (oscillations and waves in inhomogeneous
structures). The sounding by small clusters of $d$-ions cannot change
essentially the plasma parameters (although some distortion of the micro-field
in the plasma can be stimulated by the traps, which stabilize the
$d$-clusters). The advantage of static diagnostics is the simplicity of the
measurements of the inhomogeneous structure and the simple connection  with the parameters
of the interaction between $d$-ions, their shielding and the characteristics of rf
plasma. The precise theoretical consideration of the dynamical experiments
\cite{Peters,Pieper}, which are based on the excitation of the eigenmodes in
linear chains and dust crystals, seems a more complicated problem.

We would like to stress that the most general consideration of the equilibrium
inhomogeneous configurations of dusty systems can be based on
translationally non-invariant solutions of the connected system of
kinetic equations for plasmas and Poisson's equation, where the
separation between an external field and $d-d$ interaction is
absent. The equilibrium positions for the $d$-ions can be found as the points of space
where the self-consistent electric field is in balance with gravity. However,
this program is too complicated and, as we showed, not necessary for a
reasonable theoretical description of the existing experiments.

We are grateful to Dr. A.Melzer for fruitful discussions and private
communications about the experimental results. We also would like to thank
Dr. H.Thomas and Dr. J.Goree for private communications connected with
the papers \cite{Morfill,Thom,PiGoQu}.

This work have been performed with the support of INTAS grant N 96-0617.

\newpage

\begin{figure}
\caption{An example of inhomogeneous plane dusty crystals (2D crystal).}
\label{fig1}
\end{figure}

\begin{figure}
\caption{Linear chain of dust particles (number of particles $N=12$) in
  a parabolic trap. The inhomogeneity was calculated on basis of
 Eq. (15) for $t=0.18$ and compared with the experimental
 data [1].}
\label{fig2}
\end{figure}


\begin{references}
\bibitem{Peters}  S. Peters, A. Homman, A.Melzer, A. Piel, Phys.Lett. {\bf A
223}, 389, 1996.

\bibitem{Homm}  A. Homman, A.Melzer, S. Peters, A. Piel, Phys.Rev. {\bf E 56}
7138, 1997.

\bibitem{Pieper}  J.B. Pieper, J.Goree, Phys.Rev.Lett. {\bf 77}, 3137, 1996.

\bibitem{Somm}  T.J. Sommerer, W.N.G. Hitchen, R.E.P. Harvey, J.E. Lawrer
Phys.Rev. {\bf A 43,} 4452, 1991.

\bibitem{Melzer}  A. Melzer, V.A. Schweigert, L.V. Schweigert, A. Homman, S.
Peters, A. Piel, Phys.Rev. {\bf E 54,} R46, 1996.

\bibitem{Schw}  V.A. Schweigert, L.V. Schweigert A.Melzer, A. Homman, A.
Piel, Phys.Rev. {\bf E 54,} 4155, 1996.

\bibitem{Goree}  J. Goree, private communication.

\bibitem{Melzer2}  A. Melzer, private communication.

\bibitem{Morfill}  G.E. Morfill, H. Thomas, J.Vac. Sci. Technol. {\bf A14},
490, 1996.

\bibitem{Thom}  H. Thomas, G.E. Morfill Nature, {\bf 379}, 806, 1996.

\bibitem{PiGoQu}  J.B. Pieper, J. Goree, R.A. Quinn, Phys.Rev. {\bf E 54,}
5636, 1996.

\end{references}
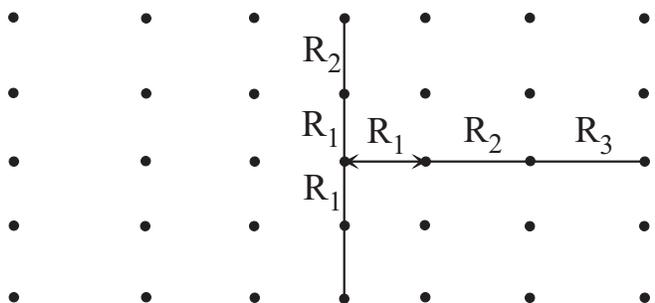
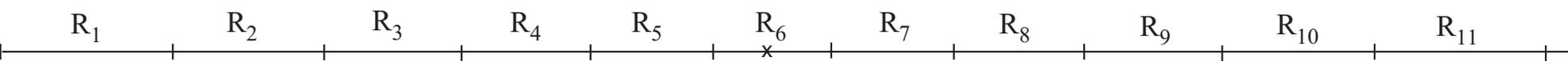
\end{document}